\documentclass[12pt]{article}

\usepackage{url}

\usepackage{natbib}  

\usepackage{graphicx}  

\usepackage{hyperref}
\providecommand{\eprint}[1]{\href{http://arxiv.org/abs/#1}{{\tt [arXiv:#1]}}}

\providecommand\apj{ApJ}                 
\providecommand\apjs{ApJSupp}                 
\providecommand\aap{A\&A}            
\providecommand\mnras{MNRAS}

\newcommand\SSS{Sect.~}

\providecommand\apj{ApJ}                 
\providecommand\apjl{ApJL}                 
\providecommand\apjs{ApJSupp}                 
\providecommand\aap{A\&A}            
\providecommand\mnras{MNRAS}
\providecommand\cqg{CQG}
\providecommand\grg{Gen. Rev. Grav.} 

\providecommand\prd{Phys.~Rev.~D}

\providecommand\jcap{JCAP}

\providecommand\grg{Gen. Rel. Grav.}

\providecommand\ISBN{ISBN:~}

\newcommand\mycaptionfont{}

\newcommand\fvir{f_{\mathrm{vir}}}
\newcommand\deltavir{\Delta_{\mathrm{vir}}}

\providecommand\agt{\,\lower.6ex\hbox{$\buildrel >\over \sim$} \, }
\providecommand\alt{\,\lower.6ex\hbox{$\buildrel <\over \sim$} \, }

\newcommand\hMpc{{$h^{-1}$~Mpc}}

\newcommand\Omm{\Omega_{\mathrm{m}}}

\newcommand\Ommeff{\Omega_{\mathrm{m}}^{\mathrm{eff}}}

\newcommand\Heff{H^{\mathrm{eff}}}

\newcommand\Omk{\Omega_{\mathrm{k}}}

\newcommand\supbg{^{\mathrm{bg}}}

\newcommand{\CD}{{\cal D}}
\newcommand{\CE}{{\cal E}}
\newcommand{\CF}{{\cal F}}
\newcommand{\CQ}{{\cal Q}}
\newcommand{\CR}{{\cal R}}
\newcommand{\CM}{{\cal M}}

\title{Simplicity in cosmology: add virialisation, remove $\Lambda$,
  keep classical GR}

\author{Boudewijn F. Roukema\\
  Toru\'n Centre for Astronomy\\
  Faculty  of Physics, Astronomy and Informatics\\
  Nicolaus Copernicus University\\
  ul. Gagarina 11\\
  87-100 Toru\'n, Poland} 

\date{le 14 juillet 2014\\
  {\em chapter to appear in\\``Mathematical structures of the Universe''}}

\begin{document}

\maketitle


\begin{abstract}
\sloppy 
Present-day extragalactic observations are mostly rather well-modelled
by a general-relativistic model, the $\Lambda$CDM model. The model
appears to surpass the limits of known physics by requiring that the
Universe be dominated by ``dark energy''.
However, the model sacrifices physical simplicity in favour
of applied mathematical simplicity. A physically simpler,
general-relativistic alternative to the $\Lambda$CDM model is described
here, along with preliminary observational checks. Thus, it will be
argued that extragalactic observations such as the
distance-modulus--redshift relation of type Ia supernovae are
well-modelled within classical general relativity, without the
addition of ``new physics''.
\end{abstract}

\newcommand\fdomaindefn{
  \begin{figure}  
    \centering
    \includegraphics[width=\textwidth]{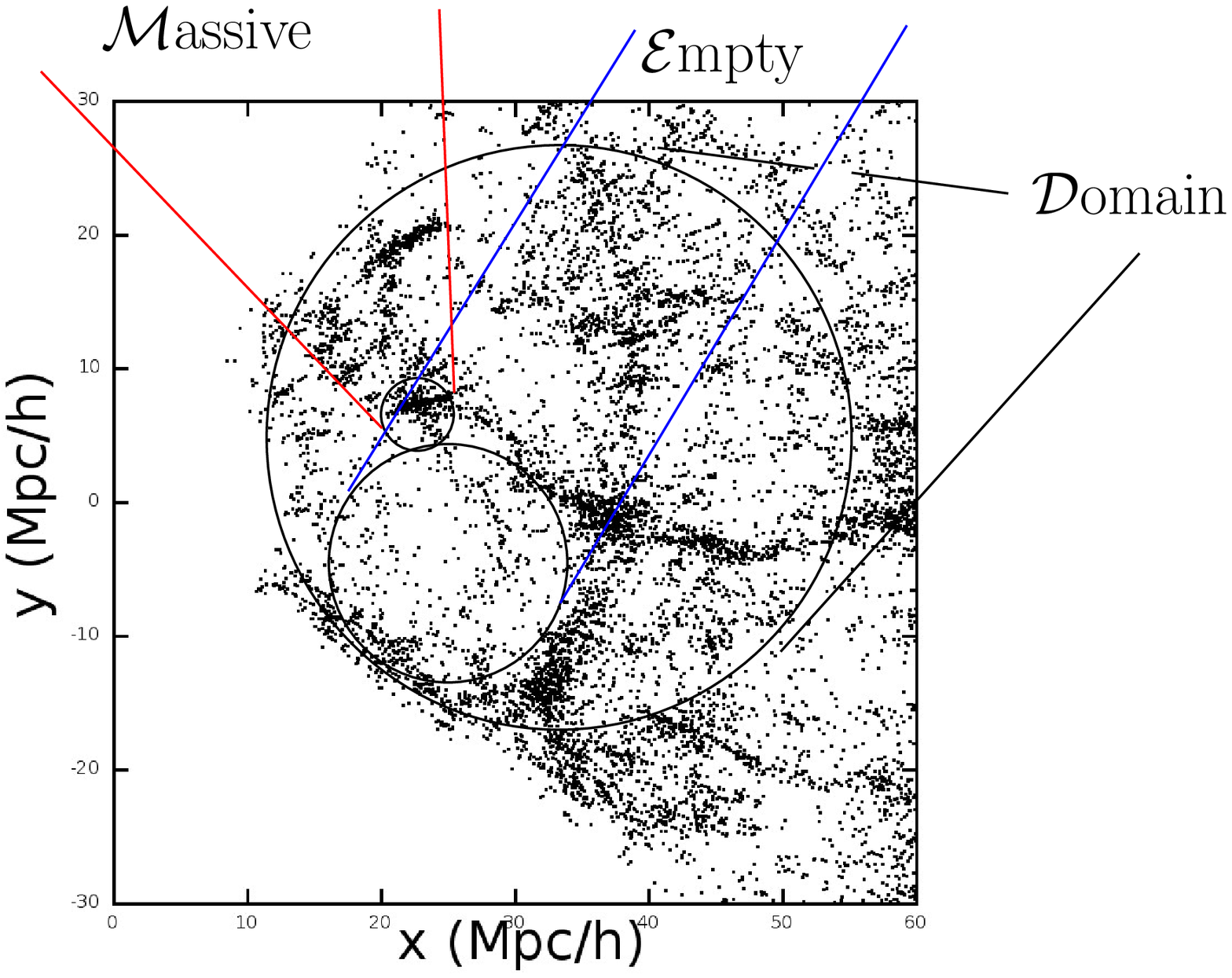}
    \caption[]{ \mycaptionfont
      Examples of ${\CM}$assive (virialised), ${\CE}$mpty (un-virialised),
      and full averaged ${\CD}$omains in a
      60{\hMpc}$\times$60{\hMpc} region showing
      galaxies observed in the two-degree--field galaxy redshift survey 
      \protect\citep[][2dFGRS, 
      \protect{\url{http://www2.aao.gov.au/2dfgrs/}}]{Coll03} 
      with the observer at the left.
      The slice is about 10 great circle degrees in thickness.
      \label{f-domaindefn}
    }
  \end{figure}
} 

\sloppy 
Within the family of locally homogeneous and isotropic solutions of
the Einstein equation, i.e. the
Friedmann--Lema\^{\i}tre--Robertson--Walker (FLRW) model
\citep{deSitt17,Fried23,Fried24,Lemaitre31ell,Rob35}, observations
since the early 1990's---faint galaxy counts and correlation functions
\citep[e.g.][]{FYTY90,1993ApJ...418L...1R,YP95}, gravitational lensing
\citep[e.g.][]{ChY97,FortMD97}, supernovae type Ia magnitude--redshift
relations \citep[e.g.][]{SCP9812,SNeSchmidt98}, and the cosmic
microwave background (CMB) from the Wilkinson Microwave Anisotropy
Probe (WMAP) \citep{WMAPSpergel} and Planck Surveyor
\citep{PlanckXVIcosmoparam13}---have required the addition of 
a parameter that does not correspond to any 
empirically detected physical phenomenon: the 
cosmological constant or dark energy parameter $\Lambda$
\citep{CosConcord95}.
This ``discovery'' has stimulated much theoretical interest,
including hypotheses of new physical components of the Universe
and theories of gravity that extend beyond classical general
relativity.

However, the FLRW solutions 
(of which the $\Lambda$CDM model is a special case)
do not take into account the fact that we
live during the virialisation epoch, i.e. the epoch during which dense
structures---galaxies, galaxy clusters, and the cosmic web
\citep[e.g.][]{deLappGH86} in general---and underdense regions---voids
on scales of many {\hMpc}---have recently formed. Thus, at small
scales and recent epochs, it should be expected that interpretations
of observational data inferred by assuming the FLRW family of models
may fail. Wrong assumptions tend to imply wrong conclusions.

Defining the fraction (by mass) of non-relativistic matter (baryonic
and dark) contained in virialised objects of typical galaxy scales and
above $\fvir(z)$ at any given redshift, it was shown in Fig.~1 of
\citet{ROB13} that the evolution of the dark energy parameter
$\Omega_\Lambda(z)$ follows that of $\fvir(z)$ to within a factor of a
few.  Thus, as the Universe evolves from high redshift (early epochs)
to low redshift (recent epochs), the virialisation fraction grows from
very little to a high fraction of unity, and the dark energy parameter
inferred from observations by {\em assuming homogeneity despite its
  increasing invalidity} grows similarly from very little to nearly
unity.  In other words, the more that the Universe becomes
inhomogeneous, the more that the assumption of homogeneity leads to
the sudden emergence of dark energy. This quantitative similarity,
$\Omega_\Lambda(z) \sim \fvir(z)$, reverses the onus of proof for dark
energy: unless or until relativistically acceptable cosmological
models including structure formation show that dark energy is needed
in order to fit observational data, the simplest explanation 
for dark energy is that it is an artefact of inhomogeneity.

Exact relativistic inhomogeneous cosmological solutions of the
Einstein equation have been known since the 1930's
\citep{Lemaitre33,Tolman34,Bondi47} and reviewed during the last few
decades \citep{Krasinski97book,Krasinski06book}.  While not directly
applicable as cosmological models without sacrificing the
Copernican principle, these solutions provide qualitative
understanding of more realistic alternatives to the FLRW model,
and can be used to model the ``holes'' in ``Swiss cheese''
cosmological models \citep[e.g.][and refs therein]{LRasSzybka13}.

\fdomaindefn

A relativistic approach to inhomogeneous cosmology that allows 
the inclusion of standard statistical representations of structure in the
Universe is the scalar averaging approach to cosmology
\citep{BKS00,Buchert08status,BuchCarf08,BuchRZA1,BuchRZA2,Kolb05a,KolbMNR05,Rasanen06superhoriz,Rasanen06GRF,Kolb11FOCUS,Wiltshire07clocks,Wiltshire07exact,Wiltshire09timescape,DuleyWilt13}. As in the FLRW approach, a spacetime
foliation is chosen, but instead of assuming homogeneity on the slices,
volume-weighted means (using the metric to measure volume) are calculated
within spatial domains of interest. 

Here we briefly describe
the virialisation approximation presented in full in 
\citet{ROB13}. This 
implementation of multi-scale scalar averaging
uses 
the definitions, derivations and terminology 
introduced in \citet{BuchCarf08} and \citet{WiegBuch10}.
Figure~\ref{f-domaindefn} illustrates the terminology,
with complementary
${\CM}$assive (virialised) and ${\CE}$mpty (un-virialised) domains,
and the full averaged ${\CD}$omains, labelled generically as ${\CF}$.
Let us define the virialisation {\em volume} fraction
\begin{equation}
  \lambda_{\CM}:=
  \frac{\left|\CM\right|}{\left|\CD\right|},
\end{equation}
which is related to $\fvir(z)$ by
\begin{equation}
  \lambda_{\CM} {=} f_{\rm vir}/\deltavir,
\end{equation}
where $\deltavir \sim 100$ to 200 
is the overdensity ratio after collapse,
e.g. estimated for a top-hat initial overdensity using
the scalar virial theorem for an isolated system
\citep[e.g.][]{LaceyCole93MN}. The homogeneous Friedmann equation
\begin{equation}
 \Omm + \Omega_\Lambda + \Omk = 1
 \label{e-ham-FLRW}
\end{equation}
can then be generalised to 
\begin{equation}
 \Omm^{\CF} + \Omega_\Lambda^{\CF} + \Omega_{\CR}^{\CF} + \Omega_{\CQ}^{\CF} = 
 \frac{H_{\CF}^{2}}{H_{\CD}^{2}}\;,
 \label{e-ham-general}
\end{equation}
where the $\CF$ superscript (or subscript) indicates which domain is being used
for averaging, the scalar-averaged expansion rate $H_{\CF}$
is defined 
\begin{equation}
  H_{\CF} := \frac{\dot{a}_{\CF}}{a_{\CF}},
\end{equation}
the rigid curvature parameter $\Omk$ has been replaced by a spatially
varying (domain-averaged) 3-Ricci curvature parameter
$\Omega_{\CR}^{\CF}$, and the kinematical backreaction parameter
$\Omega_{\CQ}^{\CF}$ arises, representing statistics (variance, shear,
and vorticity) of the extrinsic curvature tensor (in Newtonian
thinking, the velocity gradient).
See \citet{ROB13} for detailed definitions.

Considering the virialised and un-virialised regions together, i.e. 
$\CF = \CD$, the right-hand side of Eq.~(\ref{e-ham-general}) simplifies
to unity. In addition, 
approximating $|\Omega_{\CQ}^{\CF}| \ll |\Omega_{\CR}^{\CF}|$,
as found with the relativistic Zel'dovich approximation 
\citep{BuchRZA1,BuchRZA2}, the resemblance of 
Eq.~(\ref{e-ham-general}) to the rigid comoving space version,
Eq.~(\ref{e-ham-FLRW}), is obvious.
The parameter required by the FLRW model for fitting observations 
is now removed, i.e. we set $\Omega_\Lambda^{\CF} = 0$. 
For future observations, it might eventually be necessary
to allow a dark energy parameter again, though the history
of relativistic cosmology suggests that $\Lambda$ is a parameter
that falls in or out of cosmological fashion every few decades.
Thus, Eq.~(\ref{e-ham-general}) becomes
\begin{equation}
 \Omm^{\CF}  + \Omega_{\CR}^{\CF} = \frac{H_{\CF}^{2}}{H_{\CD}^{2}}\;,
 \label{e-ham-general-ROB13}
\end{equation}
or
\begin{equation}
 \Omm^{\CD}  + \Omega_{\CR}^{\CD} = 1
 \label{e-ham-general-ROB13-D}
\end{equation}
for the full domain.

To evaluate these equations, we assume the stable clustering
hypothesis \cite{Peebles1980}, i.e. 
\begin{equation}
  H_{\CM} \approx 0,
  \label{e-H-stableclustering}
\end{equation}
virialised
regions ($\CM$) are stable, we assume an Einstein--de~Sitter (EdS)
model at early epochs (high redshift), 
and we extrapolate this as a ``background''
model (with parameters labelled ``bg'')
to recent epochs (low redshift). Simplifying this set of
equations leads to Eq.~(2.22) of \citet{ROB13}, i.e.
\begin{equation}
  \Ommeff(z) := 
  \Omm^{\CD} \approx \frac{ \Omm\supbg }{
    \left({\Heff}/{H\supbg}\right)^2} , 
  \label{e-ommeff-expanded}
\end{equation}
where the label $\CD$ has been replaced by ``eff'' for convenience.

Since virialised regions are assumed to be stable
[Eq.~(\ref{e-H-stableclustering})], and since voids have to expand
faster than the extrapolated background model in order to be able
to become (nearly) empty, it is already clear that at low redshifts,
$\Heff$ will increase to become greater than $H\supbg$.
In Eq.~(2.23) of \citet{ROB13}, we model this with a smooth transition from
the pre-virialisation epoch to the present epoch, we require that
the effective expansion rate at the present epoch matches the
observed Hubble constant [Eq.~(2.32), \citet{ROB13}],
and we estimate the peculiar expansion rate of voids using void and
cluster surveys of approximately similar sizes 
[Eq.~(2.36), \citet{ROB13}]. An effective metric can then be defined,
allowing calculation of effective luminosity distances 
[{\SSS}2.4, \citet{ROB13}].

Thus, making minimal assumptions beyond the homogeneous model apart
from allowing inhomogeneous structure on spatial slices and
corresponding curvature, and defining averages on spatial domains
rather than forcing uniformity, the early-time Einstein--de~Sitter
model evolves to a model with a low matter density at the present,
$\Ommeff(0) \approx 0.3$ \citep[Fig.~2,][]{ROB13}.  Moreover, the
distance-modulus--redshift relation in the EdS+virialisation
approximation is similar to that provided by the $\Lambda$CDM model
\citep[Fig.~5,][]{ROB13}.
In other words, extragalactic observations such as the
distance-modulus--redshift relation of type Ia supernovae do not
(yet) require any extension to the present limits of physics.


\end{document}